\documentclass[doublecol]{epl2}
\usepackage{epsfig,color}
\usepackage{xcolor}
\usepackage{ulem}
\def\ADD#1{{\textcolor{black}{#1}}}    

\title{Energy cascade in internal wave attractors}

\author{C.~Brouzet\inst{1} \and E.V.~Ermanyuk\inst{1,2} \and S.~Joubaud\inst{1} \and I.~Sibgatullin\inst{1,3,4} \and T.~Dauxois\inst{1}}
\shortauthor{C.  Brouzet \etal}

\institute{
\inst{1}
 Univ Lyon, ENS de Lyon, Univ Claude Bernard, CNRS, Laboratoire de Physique, F-69342 Lyon, France
 \\
\inst{2} Lavrentyev Institute of Hydrodynamics, Novosibirsk, Russia\\
\inst{3} Institute of Mechanics and Department of Mechanics and Mathematics, Moscow State University, Russia\\
\inst{4} Shirshov Institute of Oceanology, Moscow, Russia}

\pacs{47.35.Bb}{Gravity waves}
\pacs{47.55.Hd}{Stratified flows}
\pacs{92.05.Bc}{Analytical modeling and laboratory experiments}
\pacs{47.20.-k}{Flow instabilities}

\abstract{One of the pivotal questions in the dynamics of the oceans is related to the cascade of mechanical energy in the abyss and its contribution to mixing. Here, we propose internal wave attractors in the large amplitude regime as a unique self-consistent experimental and numerical setup that models a cascade of triadic interactions transferring energy from large-scale monochromatic input to multi-scale internal wave motion. We also provide signatures of a discrete wave turbulence framework for internal waves.  Finally, we show how beyond this regime,  we have a clear transition to a regime  of small-scale high-vorticity events which induce mixing.}
\begin{document}
\def\th{\theta}
\def\l{\label}
\def\be{\begin{equation}}
\def\ee{\end{equation}}
\def\bea{\begin{eqnarray}}
\def\eea{\end{eqnarray}}
\def\b{\beta}
\def\fr{\frac}
\maketitle

\section{Introduction}

The continuous energy input to the ocean interior comes from the interaction of global tides with the bottom topography~\cite{GarrettKunze2007}. The subsequent mechanical energy cascade to small-scale internal-wave motion and mixing is a subject of active debate~\cite{IveyWintersKoseff2008} in view of the important role played by abyssal mixing in existing models of ocean dynamics~\cite{Munk1966,MunkWunsch1998,
NikurashinVallis2012}. 
A question remains: how does energy injected through internal waves at large vertical  scales~\cite{Muller1981} induce the mixing of the fluid~\cite{IveyWintersKoseff2008}?

In a stratified fluid with an initially constant buoyancy frequency $N=[(-g/\bar{\rho})({\rm d}\rho/{\rm d}z)]^{1/2}$, where $\rho(z)$ is the density distribution ($\bar{\rho}$ a reference value) over vertical coordinate~$z$, and $g$  the gravity acceleration, the dispersion relation of internal waves  is $\theta=\pm\arcsin(\Omega)$.  
The angle  $\theta$ is the slope of the wave beam to the horizontal, and
 $\Omega$  the frequency of oscillations non-dimensionalized by~$N$. The 
  dispersion relation requires preservation of the slope of the internal wave beam upon reflection at a rigid boundary. In the case of a sloping boundary, this property gives a purely geometric reason for a strong variation of the width of internal wave beams (focusing or defocusing) upon reflection. Internal wave focusing provides a necessary condition for large shear and overturning, as well as shear and bottom layer instabilities at slopes~\cite{BuhlerMuller2007,DauxoisYoung1999,ZhaKin08, GayenSarkar2010}.

In a confined fluid domain, focusing usually prevails, leading to a concentration of wave energy on a closed loop, the internal wave attractor~\cite{MBSL1997}. Attractors eventually reach a quasi-steady state where dissipation is in balance with energy injection regardless \ADD{of} the  linear~\cite{HBDM2008} or nonlinear mechanism of dissipation~\cite{JouveOgilvie2014}. High concentration of energy at attractors make them prone~\cite{Scolan2013}
 to \ADD{triadic resonance instability (TRI)}
 \ADD{, an instability similar to parametric subharmonic instability (PSI) but where viscosity plays a role}
~\cite{KoudellaStaquet2006,BDJO2013,Sutherland2013}. The resonance occurs when temporal and spatial conditions are satisfied: $\Omega_{1}+\Omega_{2}=\Omega_{0}$ and $\overrightarrow{k_{1}}+\overrightarrow{k_{2}}=\overrightarrow{k_{0}}$, where $\overrightarrow{k}$ is the wave vector while subscripts 0, 1 and 2 refer to the primary, and two secondary waves, respectively. The secondary waves can also be unstable, initiating a cascade.

In this Letter, using laboratory experiments and numerical simulations, we suggest the energy cascade  in internal wave attractors  as a novel laboratory model of a natural cascade. The cascade operates via a hierarchy of triadic interactions inducing high-vorticity events and mixing at sufficiently large forcing.
The model setup represents a trapezoidal fluid domain filled with an uniformly stratified fluid where the energy is injected at global scale by wave-like motion of the vertical wall.

Transition to mixing is non-trivial since it is clearly beyond the domain of pure wave-wave interactions. Similarly, for surface waves, experimental reality deals with the cascades of wave-wave interactions, often called wave turbulence~\cite{Nazarenko2011}, significantly `contaminated' by effects of a finite size fluid domain, wave breaking, wave cusps, nonlinear dispersion, viscous damping of wave-field components
~\cite{Nazarenko2011,Denissenkoetal2007,AubourgMordant,DeikeFalcon}. The very specific dispersion relation for internal waves introduces additional complications. For instance, in rotating fluids, which have a dispersion relation analogous to stratified fluids, the usefulness of the
 formalism of wave turbulence as a basis for the studies in rotating turbulence has been reported for experiments only recently~\cite{YaromSharon}. For internal waves, the question received some attention theoretically~\cite{TabakLvov} but remains fully open, from experimental and numerical points of view. Its consequences on mixing are moreover widely open.

\section{Experimental and numerical set-ups}
\begin{figure}[htb]
  \begin{center}
  \includegraphics[width=\linewidth,clip=]{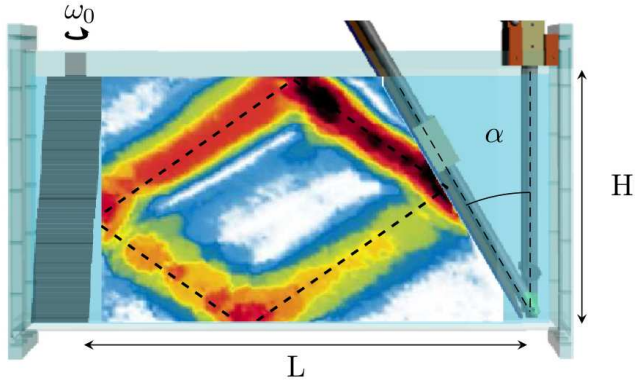}
   \end{center}
  \vspace{-0.5cm}
   \caption{Experimental set-up showing the wave generator on the left and the inclined slope on the right. The color inset is a typical PIV snapshot showing the magnitude $(u^2+w^2)^{1/2}$  of the experimental  two-dimensional velocity field obtained at $t= 15~T_0$ (case~B of Table~\ref{tabular:parameters}) with $T_0=2\pi/( N\Omega_0)$. Black dashed lines show the billiard geometric prediction of the attractor.
   \label{NumericalSimulation}}
\end{figure}

A rectangular test tank of size $80\times 17\times 42.5 ~$cm$^{3}$ is filled with a salt-stratified fluid~\cite{Scolan2013}
with $N\simeq 1$ rad$\cdot$s$^{-1}$.
 A sliding sloping wall, inclined at an angle $\alpha$ to the vertical,  delimits a trapezoidal fluid domain of length~$L$ (measured along the bottom) and depth~$H$.
The input forcing (see Fig.~\ref{NumericalSimulation}) is introduced into the system by an internal wave generator~\cite{MMMGPD2010} (left wall)
 with a time-dependent vertical profile given by
$ a\sin(N\Omega_{0} t) \cos(\pi z/H),$  where $a$  is the amplitude of oscillations.
The horizontal and vertical components of the velocity field $u$ and $w$ measured in the vertical mid-plane are then monitored as a function of spatial coordinates and time, using standard PIV technique~\cite{Westerweel1997,FinchamDelerce2000} with a cross-correlation algorithm applied to analyzing windows of typical size 20 by 20 pixels. Below, we discuss mainly three different experiments from low to large forcing.

Numerical computations are performed with spectral element methods~\cite{Fischer1997,Fischeretal}. The geometry of the numerical setup closely reproduces the experimental one. The full system of equations being solved consists of the Navier-Stokes equation in the Boussinesq approximation, the continuity equation and the equation for the transport of salt. Typical meshes used in calculations consist of 50 thousands to half-million elements, with 8 to 10-order polynomial decomposition within each element.
Time discretization was $10^{-4}$ to $10^{-5}$ of the external forcing period.
Comparisons of experimental and numerical results present a beautiful agreement, not only qualitative but also quantitative\cite{BrouzetJFM2015}.
The numerical simulations clearly emphasize the importance of \ADD{the three-dimensionality} (case Z) to recover 
 experimental laboratory results quantitatively, nevertheless 2D simulations (case X) are fully sufficient for qualitative agreement. We checked as in~\cite{Scolan2013,BrouzetJFM2015}, that 
the temporal and spatial resonance conditions are satisfied experimentally and numerically.

\section{Energy cascade revealed by the time-frequency analysis}
\begin{figure}[htb]
\begin{center}
  \includegraphics[width=0.8\linewidth,clip=]{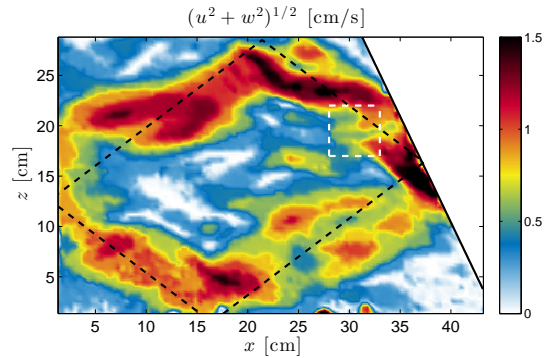}
  \end{center}
\vspace{-.5cm}
   \caption{\textit{Well developed instability}.
 Magnitude of the experimental  two-dimensional velocity field for case~B (see Table~\ref{tabular:parameters}) at $t=400\,T_0$.
 Black dashed lines show the billiard geometric prediction of the attractor, which is fully recovered when considering small forcing amplitude~\cite{Scolan2013} or at an earlier time when considering larger forcing as in Fig.~\ref{NumericalSimulation}(a). 
   \label{setup}}
\end{figure}

  An example of an experimental velocity field is shown in Fig.~\ref{setup} at a much later stage. 
  The attractor is still visible, but branches are deformed by the presence of secondary waves. As it will be clear below, the internal wave frequency spectrum which was initially a Dirac function has been progressively enriched to give
rise to a very complex spectrum, through a cascade of central interest.

\begin{figure*}[htb]
\begin{center}
 \includegraphics[width=0.8\linewidth,clip=]{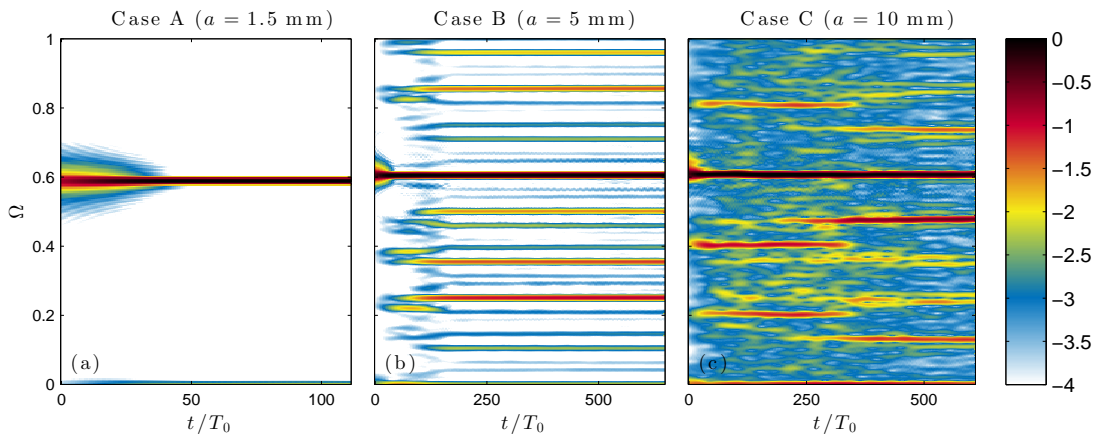}
\end{center}
\vskip -0.6truecm
\caption{\textit{Time-frequency diagrams} $\log_{10} \left(S_u(\Omega, t)/S_0\right)$, defined in Eq.~(\ref{defdeSu}), for  three different forcing amplitudes $a$. The quantity $S_0$ is
defined as the time average of the main component $S_u(\Omega_0, t)$.  The time-frequency diagrams are calculated on the $5\times5$~cm$^2$ square region, close to the reflection of the attractor on the slope and indicated by the  white dashed line in Fig.~\ref{setup}. }\label{tps_freq}
\end{figure*}

The experimental or numerical measured velocity fields are analyzed 
 using a time-frequency representation~\cite{Flandrin1999}
calculated at each spatial point.
More precisely, we compute the quantity
\begin{equation}
S_u(\Omega,t)=\left\langle \middle| \int_{-\infty}^{+\infty} \! u(x,z,\tau)e^{i\Omega\tau}h(t-\tau)\, d\tau \middle|^2\right\rangle_{xz},\label{defdeSu}
\end{equation}
where $u$ stands for the horizontal velocity component defined by the spatial coordinates $x$ and $z$,
while $h$ is a smoothing Hamming window of energy unity. The calculations are performed with the dedicated Matlab toolbox~\cite{Flandrin1999}.
To increase the signal to noise ratio, the data are averaged over the square represented in Fig.~\ref{setup}
 by the white dashed line: this is the meaning of the notation $\left\langle.\right\rangle_{xz}$. 
 We present only the analysis of the horizontal velocity field, but the results are similar for the vertical one.

Figure~\ref{tps_freq} presents
  the basic types of the newly observed cascades, with progressively increasing complexity: a simply monochromatic spectrum (case~A of Table~\ref{tabular:parameters}),  and  rich multi-peak spectra (cases {B} and~{C}). 

\begin{table}[htb]
\begin{tabular}{l|c|c|c|c|c|c|c}
 & Type & $\Omega_0$  & $H$ & $L$ & $\alpha $ & $a$ & {$t_{max}$}\\
 &  &   & cm & cm & $^{\circ}$ & mm &{$T_0$}\\
 \hline
 \hline
A&Exp. & $0.59$ & $30.0$ & $45.0$ & $27.3$ & $1.5$ & {149} \\
B&Exp.  & $0.61$ & $30.3$ & $44.4$ & $25.4$ & $5$ & {693}\\
C&Exp.  & $0.60$ & $30.1$ & $44.2$ & $24.8$ & $10$ & {651}\\
 \hline
X&2D  sim.     & 0.62 & 30.8  & 45.6 & 29.9 &  1--9 & {$1000$}\\
Z& 3D sim.  & 0.59 & 30.0  & 45.6 & 29.9 &  2.5 & {270}\\
\end{tabular}
\caption{\textit{Parameters} used for data presented in this Letter.}
\label{tabular:parameters}
\end{table}

The appropriate choice~\cite{Hamming} of the length of the Hamming window $h$ allows us to tune the resolution in frequency and time. {A} large (resp. small) window 
provides a high (resp. low) resolution in frequency and a weak (resp. good) resolution 
in time. In order to separate the different frequencies in the cases B and C, a good resolution in frequency is necessary. 
The three panels have been obtained with a 15 min long Hamming window ($\simeq80$ $T_0$).

The size of the Hamming window is also responsible of the wrong impression that the continuous 
spectrum can be seen right at the start of the experiment in Fig.~3(c). 
We checked that a time-frequency diagram with a shorter Hamming window
 emphasizes that the continuous spectrum does appear gradually like for the secondary frequency peaks in Fig.~3(b).
However,  with such a choice, the frequency resolution would not be sufficient to discriminate the frequencies.

\begin{figure}
 \begin{center}
  \includegraphics[width=0.7\linewidth,clip=]{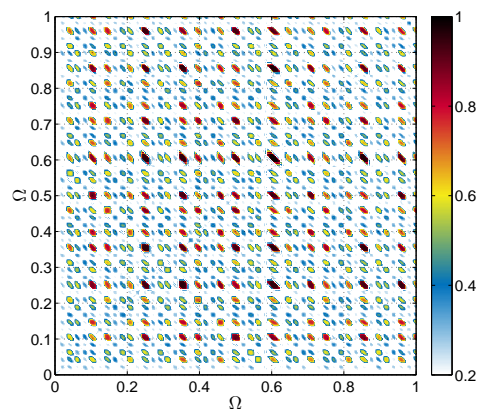} 
 \end{center}
	\vspace{-.5cm}
   \caption{Bicoherence calculated from the signal represented in Fig.~\ref{tps_freq}(b).}\label{bispectrum}
\end{figure}

In well-developed cascades,  apart from triads given by $\Omega_{i}\pm\Omega_{j}=\Omega_{0}$ associated with the primary wave oscillating at the forcing frequency $\Omega_{0}$, secondary waves are acting as primary waves for higher-order triadic interactions. Below we illustrate this by calculation of the frequency triplets for case {B} where the spectrum is rich and the discrete peaks in the spectrum are well-defined.
To detect the frequency triplets, we use the bispectrum analysis. It measures the extent of statistical
dependence among three spectral components ($\Omega_{k}$, $\Omega_{i}$, $\Omega_{j}$) 
satisfying the relationship $\Omega_{k}=\Omega_{i}+\Omega_{j}$, with the quantity
$M(\Omega_{i}, \Omega_{j})= 
F(\Omega_{i})F(\Omega_{j})F^{*}(\Omega_{i}+\Omega_{j})
$, where $F$ is the Fourier transform and $\ast$ denotes the complex conjugate. 
In practice, the bispectrum is usually normalized and considered in form of bicoherence which is $0$ for triplets with random phases and $1$ for triplets with perfect phase coupling\cite{Favier}. The bicoherence shown in Fig.~\ref{bispectrum} for case {B} is obtained using the HOSA Matlab toolbox as an average over the same square region used for the time-frequency analysis
and in the time interval $[200,690]T_0$.
 In addition to the strong peak $(0.61, 0.61)$ corresponding to the forcing frequency (therefore to self-correlation), the possible triplets satisfying the definition of triadic resonance at $\Omega_{k}=\Omega_{0}$ can be found on the line 
  with slope $-1$ connecting the points $(0, 0.61)$ and $(0.61, 0)$.
This emphasizes that the mechanism at play is triadic. Other peaks are also visible corresponding to other choices of $\Omega_{k}$ revealing that the instability mechanism is repeated and leads to a cascade.

Internal wave attractors in the large amplitude regime present therefore a nice cascade of triadic interactions transferring energy from large-scale monochromatic input to many discrete internal  wave frequencies. Moreover,
comparing Fig.~\ref{tps_freq}({b}) with Fig.~\ref{tps_freq}({c}), we note that the frequency spectrum remains qualitatively similar: a ``discrete" part,
with well-defined peaks, and a ``continuous" part (not visible in Fig.~\ref{tps_freq}({b}) but really present). However, in case~C, the magnitudes of peaks in the ``discrete" part of the spectrum fluctuate in time, and the energy content of the ``continuous" part is significantly higher (two orders of magnitude), as is clearly visible from the background color. This cascade thus presents  features reminiscent of wave turbulence, worth to explore.

\section{Signatures of discrete wave turbulence?}
\begin{figure}
 \vspace{-.5cm}
    \includegraphics[width=\linewidth,clip=]{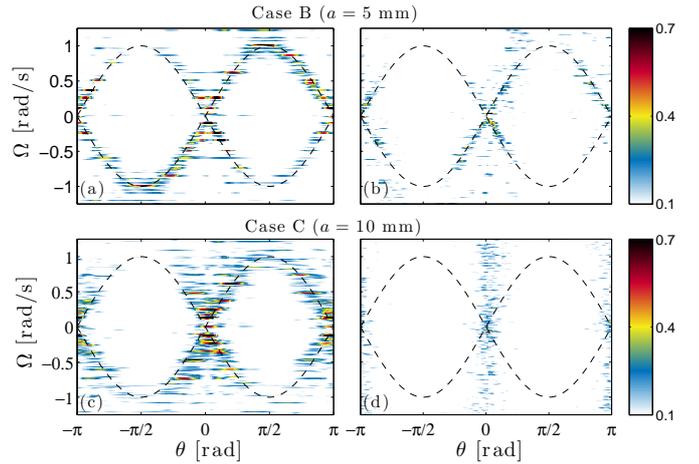}
	\vspace{-0.75cm}
   \caption{\textit{Energy spectra} \ADD{$E(\theta,\Omega)/E(\Omega)$} for two forcing amplitudes and two length scale intervals: 0.22 to 1 rad$\cdot$cm$^{-1}$ for (a) and (c);
1 to 1.86 rad$\cdot$cm$^{-1}$ for (b) and (d)  (i.e. wave lengths 28.5 cm to 6.3 cm, for left panels, while 6.3 cm to 3.4 cm for right ones). \ADD{The dashed black lines correspond to the dispersion relation $\Omega=\pm\sin\theta$.} In panels (a) and (b) which correspond to the triadic cascade experiment B,
\ADD{energy is} localized on the dispersion relation confirming the wave turbulence picture.
For the mixing box experiment C, the localization is only preserved for panel (c), while it is not
the case for panel~(d). 
 \label{WaveTUrb}}
\end{figure}

The presence of wave turbulence-like phenomena and a possible qualitative transition from discrete wave turbulence to wave turbulence-like regime with extreme events is illustrated in Fig.~\ref{WaveTUrb} using the energy spectra experimentally obtained for different wave-number intervals as a diagnostic tool~\cite{YaromSharon}. 

The wave energy spectra  are computed using the following procedure. 
Horizontal and vertical velocity fields $u(x,z,t)$ and $w(x,z,t)$ are obtained with 2D PIV measurements in the entire trapezoidal domain,  on a grid with  
0.36~cm $\times$ 0.36~cm spatial resolution and  0.5~s temporal resolution\cite{resolution}.
A three-dimensional (two dimensions for space, one for time) Fourier transform~\cite{FourierTransform} of these fields leads to 
$\hat{u}(k_x,k_z,\Omega)$ and $\hat{w}(k_x,k_z,\Omega)$. One can thus define the 2D energy spectrum by
\begin{equation}
 E(k_x,k_z,\Omega)=\frac{|\hat{u}(k_x,k_z,\Omega)|^2+|\hat{w}(k_x,k_z,\Omega)|^2}{2ST},
\end{equation}
 where $S=45\times30$ cm$^2$ is the area of the PIV measurement and $T=80~T_0$ its duration. 
 The spatio-temporal resolution of our measurements leads to upper bounds in wave numbers and frequency. We thus have $k_{\mathrm max}=8.6$~rad/cm and $N\Omega_{\mathrm max}=6.28$~rad/s.

In the dispersion relation for internal waves, $\Omega=\sin\theta$, the wave vector $\overrightarrow{k}$ and its components do not appear directly but they are linked with the angle $\theta$ by $\sin \theta=k_x/\sqrt{k_x^2+k_z^2}$. To compute the energy spectrum as a function of variable~$\theta$, one can interpolate the energy spectrum $E(k_x,k_z,\Omega)$ to get $E(k,\theta,\Omega)$, where $k$ is the norm of the wave vector. For this interpolation, we define $\Delta  k$ as the smallest wave vector that has data points in the Cartesian coordinates. Here, $\Delta k=0.043$~rad/cm and we chose $k_{\mathrm min}=5\Delta k\approx0.22$~rad/cm to have a good interpolation at low wave numbers. We chose to take $200$ points for $k$ between $0$ and $k_{\mathrm max}$ and $300$ points for $\theta$ between $-\pi$ and~$\pi$.
Then, one can integrate over the entire range of wave vectors $[k_{\mathrm min}, k_{\mathrm max}]$ as follows
 \begin{equation}
E(\theta,\Omega)=\int_{k_{\mathrm min}}^{k_{\mathrm max}}E(k,\theta,\Omega)k {\mathrm d}k,
\end{equation}
or on any range of wave vectors between $k_{min}$ and $k_{max}$. \ADD{Because the energy levels of the different frequencies cover several orders of magnitude, one has to normalize the energy density $E(\theta,\Omega)$ by the frequency energy density $E(\Omega)$, obtained by integrating $E(\theta,\Omega)$ on all $\theta$ range.}

This is what has been done in Fig.~\ref{WaveTUrb}\ADD{. The} two \ADD{spatial} integration ranges are $[0.22, 1]$ and $[1, 1.86]$~rad/cm, for cases~B and C. For case B, the first integration range represents $84\%$ of the energy in the entire range $[k_{min}, k_{max}]$ while the second  represents $11\%$. For case C, the first range has $82\%$ of the total energy and the second one $11\%$.

 The linear dispersion relation is seen to attract the maxima of the energy spectra regardless of the length scales in case B, and for large-scale perturbations only,  in case C. Short-scale perturbations in the latter case clearly escape any relation to  linear wave dynamics. This is expected to be due to extreme events, natural precursors to mixing.

Above results are convincing signatures of a discrete wave turbulence framework for internal waves in the intermediate
forcing amplitude regime. For the largest amplitude, we have  
indications that a system is beyond the wave turbulence-like regime and has reached a mixing regime.

\section{Mixing inferred from vorticity distribution}
\begin{figure}
	\begin{center}\includegraphics[width=0.9\linewidth,clip=]{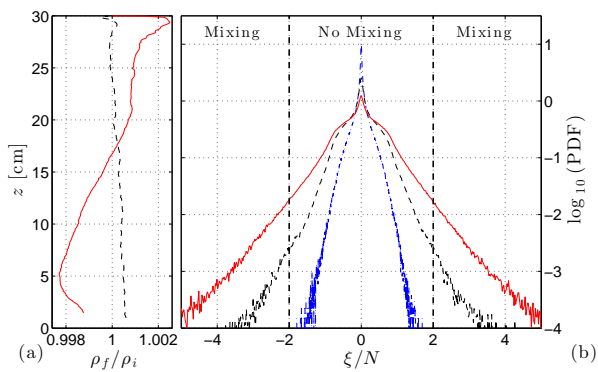}\end{center}
\vskip -0.4truecm
   \caption{\textit{Mixing and vorticity}. 
   (a) Ratio between the density profiles measured after and before the experiments for cases B (black) and  C (red).   (b) Experimental probability density functions of the vorticity  $\xi (x,z,t)$ in the tank, calculated   on the grid
   from experimental images for cases A (blue), B (black) and C (red). Samples are taken using 400 images close to the end of the experiment, when the wave regime is fully developed.
     The individual PDFs are averaged over roughly $8\times10^3$ equally spaced points covering the whole wave field, and are normalized by the buoyancy frequency~$N$. 
    \label{PDF_vorticity}}
\end{figure}

An important issue is whether or not sufficiently energetic internal wave motion can produce an irreversible energy contribution to mixing. Figure~\ref{PDF_vorticity}(a) presents
the comparison between density profiles measured before and after experiments: while no modification of the density (within experimental error)
can be observed in case B, one gets a clear evidence of mixing in case C.

Further,
 differences between the regimes corresponding to low and high mixing are clearly seen in statistics of extreme events. This statistic is obtained by the calculation of probability density functions (PDF),  a widely used tool for describing turbulence~\cite{Batchelor1982}.
  Since we are interested in small-scale events destabilizing the stratification, we take the horizontal $y$-component of vorticity $\xi (x,z,t)=\partial u/\partial z-\partial w/\partial x$ measured in the vertical midplane of the test tank as a relevant quantity and consider its PDF.  In Fig.~\ref{PDF_vorticity}(b), we present
the vorticity PDFs corresponding to different wave regimes in the attractor. Note that the area under each PDF is equal to unity, \ADD{which} allows a meaningful comparison between the probabilities of extreme events in the cases A, B and C.
 In a stable attractor (see case~A), extreme events \ADD{(defined with respect to $2N$, see below)} are completely absent and the wave motion is concentrated within the relatively narrow branches of the attractor while the rest of the fluid is quiescent. Accordingly, the PDF has a sharp peak at zero vorticity and is fully localized between well-defined maximum and minimum values of vorticity. In cases B and C, \ADD{the PDF broadens and} the development of \ADD{TRI} increases the probability of extreme events due to summation of primary and secondary wave components. 
 
The occurrence of local destabilizing events can be viewed as a competition between  stratification and vorticity. In a two-dimensional flow, a relevant stability parameter is a version of the Richardson number, which can be introduced as $\mathrm{Ri}_{\xi}=N^2/\xi^2$. For a horizontal stratified shear flow this parameter reduces to the conventional gradient Richardson number
$ \mathrm{Ri}=N^2/(\mbox{d}u/\mbox{d}z)^2,$
 where $\mbox{d}u/\mbox{d}z$ is the velocity shear. Flows with large $\mathrm{Ri}$ are generally stable, and the turbulence is suppressed by the stratification. The classic Miles-Howard necessary condition for instability requires that $\mathrm{Ri}{<}1/4$ somewhere in the flow. If this condition is satisfied, the destabilizing effect of shear overcomes the effect of  stratification, and some mixing occurs as a result of overturning. The threshold value $|\xi /N|=2$ is marked on the plot of vorticity PDFs. It can be seen that  data corresponding to cases B and C have "tails" extending into the domains  $|\xi /N|>2$. The area under the tails represents the probability of event\ADD{s} of strength $|\xi /N|>2$. In case C, this probability is an order of magnitude greater than in case B, 
 in qualitative agreement with much higher mixing in case C as compared to case~B.

 The measure of the mixing can be defined as the normalized potential energy
  $A(t)=(E_p(t)-E_p(0))/(E_p^*-E_p(0))$, in which $E_p=\int{\rm d}x{\rm d}z\, \rho(x,z,t) g z$ stands for the potential energy and $*$ stands for its final value for the fully mixed system.
  For the mixing box experiment (see profile C shown in Fig.~\ref{PDF_vorticity}(a)),
 one attains a final value $A\approx 25\%$. Mixing is therefore remarkably strong: two hours of experiment in case C are equivalent to the action of molecular diffusion on a time scale of several weeks.

\begin{figure}[htb]
\begin{center}
\includegraphics[width=0.7\linewidth,clip=]{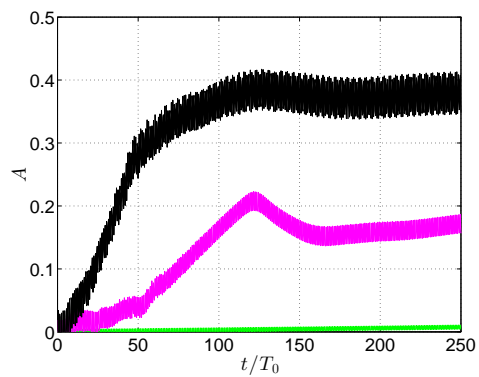}
\end{center}
\vskip -0.6truecm
   \caption{\textit{Potential energy}.
   Time evolution of the normalized potential energy for three different forcing amplitudes:  green ($a=1$ mm), magenta ($a=5$ mm) and black ($a=9$ mm).  The curves were obtained using long duration 2D numerical simulations  for case~X of Table~\ref{tabular:parameters}.
   \label{EnergyPotential}}
\end{figure}

The density profiles  measured before and after the experimental runs do not allow monitoring the time evolution of the mixing dynamics. However, these dynamics are nicely revealed in numerical calculations as shown in Fig.~\ref{EnergyPotential}. The dramatic effect of the amplitude of oscillations on the mixing (with other parameters being fixed) is clearly seen, ranging from slow erosion of initial stratification to violent mixing.

\section{Conclusions} In the present Letter, we have reported and described a novel experimental and numerical setup,  an ``internal wave mixing box", which presents a complete cascade of triadic interactions transferring energy from large-scale monochromatic input to multi-scale internal wave motion, and subsequent cascade to mixing. We have reported interesting signatures of discrete wave turbulence in a stratified idealized fluid problem. Moreover, we have shown how 
\ADD{extreme vorticity events lead to mixing that occurs} in the bulk of the fluid, similar to~\cite{SwartMandersHarlanderMaas}.
 
Confinement of the fluid domain and focusing of wave energy at an attractor play an important role in the cascade; however, these conditions are not very restrictive. Under natural conditions, internal waves can travel thousands of kilometers which means that quite large bodies of water (for instance, seas) can be considered as confined domains. Also, since attractors can occur in laterally open domains with an appropriately shaped bottom~\cite{EYBP2011}, the mechanism of the triadic wave cascade and the bulk mixing described in the present paper is likely to occur in domains with multi-ridge topography as described in~\cite{Polzin1997}.

From a broader perspective, the complete scenario that we have identified here thus provides  an analog of energy cascade in the abyss that should shed new light on the full energy cascade in the oceans. However, the quasi-two dimensional set-up and the absence of Coriolis forces hinders, at this stage, the generalization beyond
the idealized fluid problem that we study here. Work along these lines to more closely reproduce the 
cascade in the oceans would be highly interesting.

\section{Acknowledgements}
EVE gratefully acknowledges his appointment as a Marie Curie incoming fellow
at  ENS de Lyon. This work has been partially supported
by  ONLITUR grant (ANR-2011-BS04-006-01), by Russian ministry of education (RFMEFI60714X0090), RFBR (15-01-06363) and CFD web-laboratory unihub.ru. It has  been
achieved with resources of PSMN from ENS de Lyon. Most of numerical simulations were performed on supercomputer «Lomonosov» of Moscow State University.
We thank P. Borgnat, P. Flandrin, N. Mordant,
 A. Obabko, H. Scolan, A. Venaille, A. Wienkers for helpful discussions.


\begin{thebibliography}{99}
\bibitem{Hamming}
As the size of the Hamming window is on the order of the duration of the experiments, some edge problems appear 
at the beginning and end of the time-frequency diagrams. This can be clearly seen in Fig.~\ref{tps_freq}, 
at the beginning of all diagrams, around $\Omega_0$, where the peak is wider than in the rest of the experiments. This 
problem is not visible at the end because we deliberately cut off diagrams just before.
\bibitem{resolution}
2D PIV measurements are performed with a $10\times10$ pixel resolution. 
The $x$ direction length being $45$~cm and the $z$ direction length $30$~cm,
this leads to a resolution of $0.36$~cm in both directions. 
The temporal resolution is $0.5$~s and we used $40~T_0$ to compute each energy spectrum. Thus, the total data grid size is $125\times80\times800$ points.
\end{thebibliography}
\end{document}